\documentclass[structabstract]{aa} 
\usepackage{CJK}
\usepackage{booktabs}
\usepackage{amsmath}
\usepackage{graphicx}
\usepackage{txfonts}
\usepackage{natbib}
\usepackage{longtable}
\usepackage{lscape}
\usepackage{multirow}
\usepackage{pdfpages}
\usepackage{ulem}
\usepackage[colorlinks=true, allcolors=blue]{hyperref}

\bibpunct{(}{)}{;}{a}{}{,}

\newcommand{\kms}{km\,s$^{-1}$}

\newcommand{\degree}{$^{\circ}$}

\begin{document}

%
\title{Sulfur isotope ratios in the Large Magellanic Cloud}

\author{Y.~Gong\inst{1}, C.~Henkel\inst{1,2}, K.~M.~Menten\inst{1}, C.-H.~R. Chen\inst{1}, Z.~Y.~Zhang\inst{3}, Y.~T.~Yan\inst{1}, A.~Weiss\inst{1}, N.~Langer\inst{4}, J.~Z.~Wang\inst{5}, R.~Q.~Mao\inst{6,7}, X.~D.~Tang\inst{2,7,8}, W.~Yang\inst{3,1}, Y.~P.~Ao\inst{6,7}, M.~Wang\inst{6,7}}
\offprints{Y.~Gong, \email{ygong@mpifr-bonn.mpg.de}}

\institute{
Max-Planck-Institut f{\"u}r Radioastronomie, Auf dem H{\"u}gel 69, D-53121 Bonn, Germany
\and 
Xinjiang Astronomical Observatory, Chinese Academy of Sciences, 150 Science 1-Street, Urumqi, Xinjiang 830011, China
\and 
School of Astronomy \& Space Science, Nanjing University, 163 Xianlin Avenue, Nanjing 210023, People's Republic of China
\and 
Argelander-Institut für Astronomie, Universität Bonn, Auf dem Hügel 71, 53121 Bonn, Germany
\and
Guangxi Key Laboratory for Relativistic Astrophysics, Department of Physics, Guangxi University, Nanning 530004, PR China
\and 
Purple Mountain Observatory, Chinese Academy of Sciences, Nanjing 210023, PR China
\and
Key Laboratory of Radio Astronomy, Chinese Academy of Sciences, Nanjing 210023, PR China
\and 
University of Chinese Academy of Sciences, Beijing 100049, PR China
}

\date{Received date ; accepted date}

\abstract
{Sulfur isotope ratios have emerged as a promising tool for tracing stellar nucleosynthesis, quantifying stellar populations, and investigating the chemical evolution of galaxies. While they are extensively studied in the context of the Milky Way, they still remain largely unexplored in extragalactic environments.}
{We focus on investigating the sulfur isotope ratios in the Large Magellanic Cloud (LMC) to gain insights into sulfur enrichment in this nearby system and to establish benchmarks for such ratios in metal-poor galaxies.}
{We conducted pointed observations of CS and its isotopologues toward N113, one of the most prominent star-formation regions in the LMC, utilizing the Atacama Pathfinder EXperiment 12~m telescope.}
{We present the first robust detection of C$^{33}$S in the LMC by successfully identifying two C$^{33}$S transitions on a large scale of $\sim$5~pc. Our measurements result in an accurate determination of the $^{34}$S/$^{33}$S isotope ratio, which is 2.0$\pm$0.2. Our comparative analysis indicates that the $^{32}$S/$^{33}$S and $^{34}$S/$^{33}$S isotope ratios are about a factor of two lower in the LMC than in the Milky Way.
}
{Our findings suggest that the low $^{34}$S/$^{33}$S isotope ratio in the LMC can be attributed to a combination of the age effect, low metallicity, and star formation history.}


\titlerunning{First detection of C$^{33}$S in the LMC}

\authorrunning{Y. Gong et al.}

\maketitle


\section{Introduction}
Isotope abundance ratios provide salient information for tracing stellar nucleosynthesis, evaluating stellar ejecta, quantifying stellar populations, constraining chemical evolution, and unraveling the history of chemical enrichment in the Universe \citep[e.g.,][]{1994ARA&A..32..191W,2018Natur.558..260Z}.
While carbon, nitrogen, and oxygen isotopes are primarily synthesized in the CNO cycles and helium burning processes \citep[e.g.,][]{1994ARA&A..32..191W,2022A&ARv..30....7R}, sulfur isotopes provide unique insights into the chemical enrichment by oxygen-burning, neon-burning, and s-process nucleosynthesis. Core-collapse supernovae, Type Ia supernovae, novae, and asymptotic giant branch (AGB) stars are the key astrophysical sources contributing to the production of sulfur isotopes \citep[e.g.,][]{1992A&ARv...4....1W,1994ARA&A..32..191W,1998ApJ...494..680J,2020ApJ...900..179K}. Therefore, sulfur isotopes hold immense potential to fill gaps in our understanding of stellar nucleosynthesis and the chemical evolution of galaxies.


Sulfur, with its four stable isotopes, $^{32}$S,  $^{34}$S, $^{33}$S, and $^{36}$S, exhibits abundance ratios of 95.02:4.21:0.75:0.02 in the Solar System \citep{1989GeCoA..53..197A,2003ApJ...591.1220L}. Carbon monosulfide ($^{12}$C$^{32}$S, hereafter CS) is the most abundant sulfur-bearing molecule in the interstellar medium (ISM). Extensive observations of its isotopologues C$^{32}$S, C$^{34}$S, C$^{33}$S, and C$^{36}$S have provided sulfur isotope ratios across the Milky Way \citep[e.g.,][]{1996A&A...305..960C,2020ApJ...899..145Y,2020A&A...642A.222H,2023A&A...670A..98Y}. Moreover, the main isotopologue, CS, is found to be also ubiquitous in external galaxies \citep[e.g.,][]{1985A&A...150L..25H,1995AJ....109.1716P,2009ApJ...690..580W,2011MNRAS.416L..21W,2014ApJ...784L..31Z}. In contrast, its rare isotopologues are scarcely detected beyond the Milky Way. So far, extragalactic C$^{34}$S has been detected in emission only in the Large Magellanic Cloud (LMC), NGC~4945, NGC~253, M82, and likely the strongly lensed galaxy, APM 08279+5255 at a redshift, $z$, of 3.911 \citep{1989A&A...223...79M,2004A&A...422..883W,2006ApJS..164..450M,2009ApJ...690..580W,2020ApJ...891..164S,2021A&A...656A..46M,2023arXiv230807368Y}, while C$^{33}$S has been identified solely in the LMC, the Small Magellanic Cloud (SMC), and NGC~253 \citep{2018ApJ...862..102S,2020ApJ...891..164S,2021A&A...656A..46M}. In addition, absorption from C$^{34}$S (and CS) was also detected toward the lensing galaxies of the intermediate redshift, $z$, namely:\ gravitational lens systems B0218+357 and PKS 1830$-$211 at $z =0.68$ and 0.89, respectively, whose strongly magnified continuum background sources afford sensitive measurements of both species \citep{Muller2006, Wallstrom2016}.
Additionally, C$^{36}$S remains undetected outside the Milky Way. Therefore, sulfur isotope ratios in extragalactic environments remain largely unexplored.

At a distance of $\sim$50~kpc \citep[e.g.,][]{2013Natur.495...76P}, the LMC is one of the closest metal-poor galaxies with a metallicity of $\sim$0.3--0.5~$Z_{\odot}$ \citep{1997macl.book.....W}. Based on observations of the UV spectra of H{\scriptsize II} regions, it has been found that the carbon, nitrogen, oxygen, and sulfur element abundances in the LMC are a factor of about 6, 10, 3, and 2 lower than in the solar neighborhood \citep[e.g.,][]{1982ApJ...252..461D,2016ApJ...818..161N}, respectively. Given these data and its proximity, the LMC is the ideal target for a first step to investigate the isotope ratios in metal-poor galaxies. Sulfur isotopic ratios have been investigated in two hot cores, ST11 and ST16, in the LMC \citep{2020ApJ...891..164S}. However, it is important to note that these ratios were determined with a certain degree of uncertainty, which raises questions about their similarity to the Galactic values as determined by \citet{2023A&A...670A..98Y}. In addition, whether these ratios observed on the small scale of $\sim$0.1 pc can be representative of the entire LMC or not remains uncertain. Measuring these ratios on larger spatial scales in different sources can offer valuable insights to address this question.

N113 is a prominent massive star-formation region and one of the most prolific sources of molecular line emission in the LMC \citep{1997A&A...317..548C,1998A&A...332..493H,2009ApJ...690..580W,2014A&A...572A..56P,2016ApJ...818..161N}. Furthermore, N113 hosts the most intense H$_{2}$O maser in the Magellanic Clouds \citep{2010MNRAS.404..779E}, one 1665~MHz OH maser \citep{1997MNRAS.291..395B}, both signs of ongoing star formation and several young stellar objects \citep[e.g.,][]{2009ApJS..184..172G,2010A&A...518L..73S,2012A&A...542A..66C}. Given these distinctive properties coupled with the pronounced intensities of its spectral lines, N113 presents itself as an exceptional candidate for the pursuit of rare isotopologues of molecules. \citet{2009ApJ...690..580W} reported a tentative detection of C$^{33}$S (3-2) toward N113 on a large scale, but its velocity centroid of 238.38$\pm$0.53~\kms\,significantly deviates from the value of the corresponding C$^{34}$S and CS lines, $234.82\pm0.11$ and $235.13\pm0.09$ ~\kms, respectively (along with those of a plethora of lines from other molecules), which casts doubt on its reliability. Therefore, we performed dedicated observations of CS and its isotopologues toward N113 to shed light on sulfur isotope ratios on a linear scale of $\sim$5~pc in low-metallicity galaxies.

Our observations are described in Sect.~\ref{Sec:obs}. In Sect.~\ref{Sec:res}, we report our discoveries. The results are discussed in Sect.~\ref{Sec:dis}. Our summary and conclusions are presented in Sect.~\ref{Sec:sum}.

\section{Observations and data reduction}\label{Sec:obs}
In the months of  April through June 2023 (project code: M9514C\_111), pointed observations of N113 were carried out using the Atacama Pathfinder EXperiment 12 meter submillimeter telescope \citep[APEX;][]{2006A&A...454L..13G}. The nFLASH230\footnote{\url{https://www.apex-telescope.org/ns/nflash/}} and SEPIA180\footnote{\url{https://www.apex-telescope.org/ns/instruments/sepia/sepia180/}} receivers \citep{2018A&A...611A..98B,2018A&A...612A..23B} were employed to observe the $J=4-3$ and $J=5-4$ transitions of CS and its isotopologues, respectively. The nFLASH230 and SEPIA180 frequency setups offered instantaneous intermediate frequency (IF) bandwidths of 16~GHz and 8~GHz, respectively. The backend used for data processing was an evolved version of the fast Fourier transform spectrometers \citep[FFTSs;][]{2012A&A...542L...3K}, which covered the above-mentioned IF bandwidth with overlapping 4~GHz wide modules with 65,536~channels each, resulting in a channel width of 61~kHz. 
The receivers were tuned to the frequencies of 194.5, 216.1, 242, and 250 GHz in the lower sideband to observe the target lines listed in Table~\ref{Tab:lin}. 

The observations were performed in the position-switching mode using the APECS software \citep{2006A&A...454L..25M}. The telescope was pointed toward ($\alpha_{\rm J2000}$, $\delta_{\rm J2000}$)= (05$^{\rm h}$13$^{\rm m}$17\rlap{.}$^{\rm s}$40, $-$69\degree22\arcmin22\rlap{.}\arcsec0), the position of the molecular clump with the strongest HCN and HCO$^+$ ($1-0$) emission in N113 
identified by \citet[][their clump 4]{Seale2012} which is located at an offset of ($-$12\arcsec,+13\arcsec) relative to the pointing position in \citet{2009ApJ...690..580W}. The off position was ($\alpha_{\rm J2000}$, $\delta_{\rm J2000}$)=(05$^{\rm h}$14$^{\rm m}$21\rlap{.}$^{\rm s}$127, $-$69\degree23\arcmin33\rlap{.}\arcsec99).
Calibration was carried out approximately every five minutes. System temperatures ranged from 83 to 355~K on a $T_{\rm A}^{*}$ scale. Based on measurements of planets\footnote{\url{http://www.apex-telescope.org/telescope/efficiency/}}, the main beam efficiencies of 86\% and 81\% were utilized for the $J$=4--3 and $J$=5--4 transitions, respectively. These efficiencies were used to convert the intensity scale to main beam brightness temperatures for the observed transitions. The half-power beam widths (HPBWs) for the observed transitions are listed in Table~\ref{Tab:lin}. 

The data reduction was carried out using the GILDAS software\footnote{\url{https://www.iram.fr/IRAMFR/GILDAS/}} \citep{2005sf2a.conf..721P}. A linear baseline was subtracted from the spectra for subsequent analysis.

\begin{table*}[!hbt]
\caption{Observational and physical parameters of the transitions of CS and its isotopologues.}\label{Tab:lin}
\normalsize
\centering
\begin{tabular}{cccccccc}
\hline \hline
line             & Frequency         & $\theta_{\rm beam}$  & $\varv_{\rm LSR}$ & $T_{\rm p}$  & $\Delta \varv$ & $\int T_{\rm mb} {\rm d}\varv$  & $N$ \\ 
                 & (GHz)             &  (\arcsec)   & (\kms)    & (mK)   & (\kms)    & (mK~\kms) & (cm$^{-2}$)\\ 
(1)              & (2)               & (3)          & (4)       & (5)   & (6)   & (7)  & (8)  \\
\hline
CS $J=4-3$        & 195.9542109(16)       & 30 & 234.5$\pm$0.1 & 448.8$\pm$13.5 & 4.71$\pm$0.02 & 2251$\pm$8  & (1.4$\pm$0.1)$\times 10^{13}$ \\
C$^{34}$S $J=4-3$ & 192.8184566(12)       & 31 & 234.3$\pm$0.1 & 33.4$\pm$3.6   & 4.68$\pm$0.34 & 167$\pm$10  & (1.1$\pm$0.1)$\times 10^{12}$ \\
C$^{33}$S $J=4-3$ & 194.3365623(3)        & 30 & 234.3$\pm$0.2 & 15.5$\pm$1.9   & 3.83$\pm$0.68 &  85$\pm$11   & (5.3$\pm$0.8)$\times 10^{11}$ \\
CS $J=5-4$        & 244.9355565(28)       & 24 & 234.5$\pm$0.1 & 410.1$\pm$17.9 & 4.84$\pm$0.04 & 2113$\pm$13 & (1.8$\pm$0.3)$\times 10^{13}$ \\
C$^{34}$S $J=5-4$ & 241.0160892(7)        & 25 & 234.4$\pm$0.1 & 27.8$\pm$1.2 & 4.22$\pm$0.21 & 125$\pm$6     &  (1.1$\pm$0.2)$\times 10^{12}$ \\
C$^{33}$S $J=5-4$ & 242.9136103(3)        & 24 & 234.5$\pm$0.3 & 8.9$\pm$1.4 & 3.52$\pm$0.66 & 36$\pm$11       & (3.0$\pm$1.1)$\times 10^{11}$ \\
C$^{36}$S $J=5-4$ & 237.5258421(9)        & 25 & ...  & $<$5.3 & ... & $<$26  & $<$2.2$\times 10^{11}$ \\
$^{13}$CS $J=5-4$\tablefootmark{a} & 231.2206852(35)       & 26 & 234.8$\pm$1.0  &  9.0$\pm$2.6 & 5.71$\pm$2.97 & 55$\pm$21  & (4.7$\pm$2.0)$\times 10^{11}$ \\
$^{13}$C$^{34}$S $J=5-4$ & 227.3005058(35)& 26 & ...  & $<$13.6& ... & $<$69  & $<$5.7$\times 10^{11}$ \\
\hline
\end{tabular}
\tablefoot{(1) Transition. (2) Rest frequency taken from the Cologne Database for Molecular Spectroscopy \citep[CDMS,][]{2016JMoSp.327...95E}. Uncertainties in the last digits are given in parentheses. (3) Half-power beam width. (4) Velocity centroid. (5) Peak main beam brightness temperature. (6) Full width at half maximum (FWHM) line width. (7) Integrated intensity.
The integrated intensities are obtained through Gaussian fitting, with the exception of two C$^{33}$S lines, which are estimated by integrating across the velocity range of 225--245~\kms. For C$^{36}$S and $^{13}$C$^{34}$S (5-4), the 3$\sigma$ upper limits of integrated intensities are also integrated over the velocity range of 225--245~\kms. (8) Total molecular column density derived assuming Local Thermodynamic Equilibrium (LTE) conditions, optically thin emission, and an excitation temperature of 13.1$\pm$1.2~K. ``..." indicates that no information is available. \tablefoottext{a}{Tentative detection}.}
\normalsize
\end{table*}

\section{Results}\label{Sec:res}
\begin{figure*}[!htbp]
\centering
\includegraphics[width = 1.00 \textwidth]{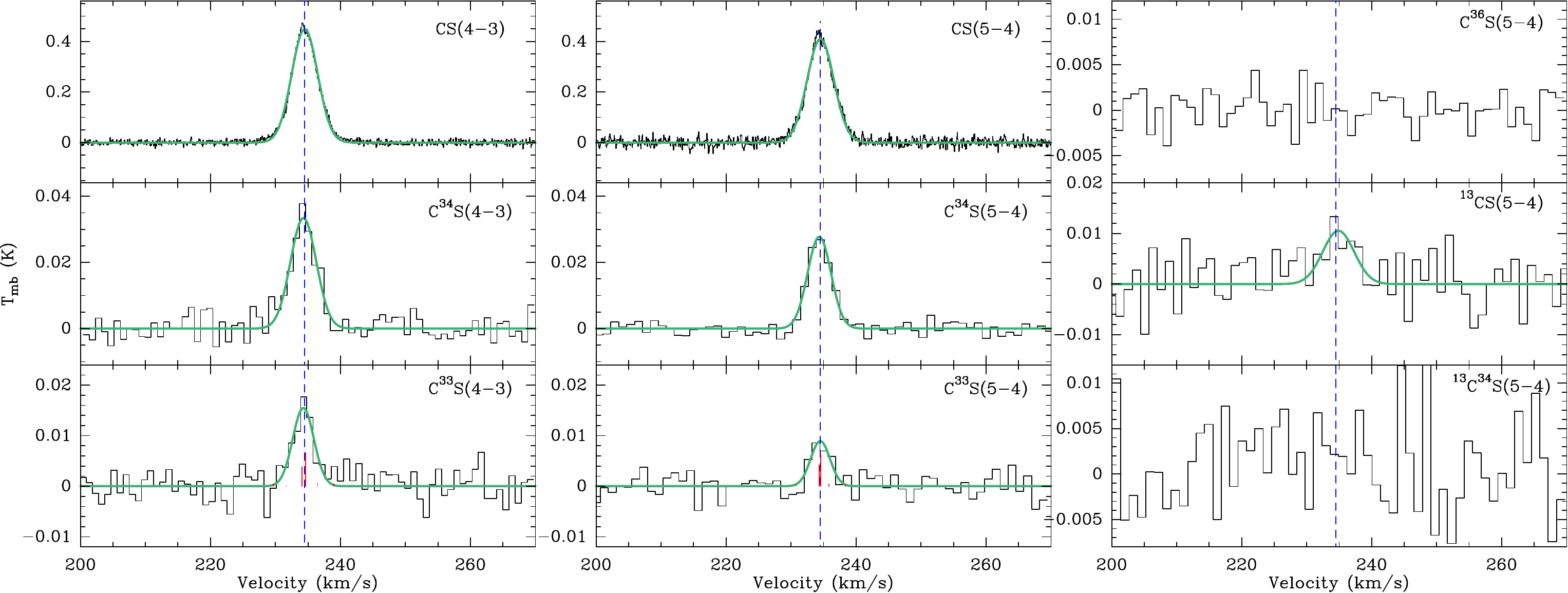}
\caption{{Observed transitions of CS and its isotopologues toward N113 overlaid with Gaussian fits indicated in green. The corresponding transition is indicated in the top-right corner of each panel. The blue dashed line indicates the systemic velocity of N113. The positions and relative intensities of the HFS components of C$^{33}$S (4-3) and C$^{33}$S (5-4) are indicated by the red vertical lines. In this plot, all spectra except CS (4-3) and CS (5-4) have been smoothed to a channel width of $\sim$1~\kms.}\label{Fig:spec}}
\end{figure*}

In Fig.~\ref{Fig:spec}, we present the spectra of CS and its isotopologues observed toward N113. Using 3$\sigma$ detection criteria, our APEX observations have led to the successful detection of the CS (4-3), C$^{34}$S (4-3), C$^{33}$S (4-3), CS (5-4), C$^{34}$S (5-4), and C$^{33}$S (5-4) lines. As the detected spectra exhibit approximately Gaussian profiles, we adopted a single Gaussian component for fitting the observed spectra, and the fitted values are given in Table~\ref{Tab:lin}. Due to the nonzero nuclear spin of $^{33}$S, C$^{33}$S transitions have hyperfine structure (HFS),  which is indicated by the red vertical lines in Fig.~\ref{Fig:spec}. Neglecting these HFS lines can potentially bias the $^{32}$S/$^{33}$S and $^{34}$S/$^{33}$S isotope ratios \citep{2023A&A...670A..98Y}. To address this concern, we integrated over the velocity range of 225--245~\kms, encompassing all HFS lines of C$^{33}$S.

Our observations also led to the tentative detection of $^{13}$CS (5-4) with a signal-to-noise ratio of 2.6 in its integrated intensity. C$^{36}$S (5-4) and $^{13}$C$^{34}$S (5-4) were also observed, but their intensities are too weak to achieve detection by our observations. Hence, assuming similar line shapes, only 3$\sigma$ upper limits are presented for these two transitions in Table~\ref{Tab:lin}.

The measured velocity centroids and FWHM line widths of detected transitions of CS and its rare isotopologues are consistent with each other within the 3$\sigma$ uncertainties. The simultaneous detection of two C$^{33}$S transitions at a consistent velocity secures our robust discovery of C$^{33}$S in N113, which represents the first detection of C$^{33}$S on a scale of $\sim$5~pc in a low-metallicity galaxy. To our knowledge, the LMC is the third extragalactic system with C$^{33}$S detection, after the SMC and NGC 253 \citep{2018ApJ...862..102S,2021A&A...656A..46M}.


In the optically thin regime, the excitation temperature can be estimated from the line ratio between the $J$=4--3 and $J$=5--4 transitions of the same molecule under the assumption of local thermodynamic equilibrium (LTE). Employing C$^{34}$S and assuming a source size of 40\arcsec\,\citep{2009ApJ...690..580W}, we derived an excitation temperature of 13.1$\pm$1.2~K. A similar excitation temperature of 10.3$\pm$2.5~K is obtained using the two C$^{33}$S transitions. Because the C$^{34}$S lines have higher signal-to-noise ratios, an excitation temperature of 13.1$\pm$1.2~K was adopted for all the CS isotopologues in this study. 

We also note that the excitation temperatures of the two CS transitions could be slightly different from those of C$^{34}$S and C$^{33}$S transitions. To better ascertain the excitation temperature of CS transitions, we ran the non-LTE RADEX\footnote{\url{https://var.sron.nl/radex/radex.php}} code \citep{2007A&A...468..627V}, where the molecular data of CS are directly taken from the Leiden Atomic and Molecular Database \citep[LAMDA;][]{2005A&A...432..369S} and the ortho-to-para ratio of H$_{2}$ is assumed to be 0.25 \citep{2006ApJ...649..816N}. Based on previous studies \citep[see Table~3 in][]{2017A&A...600A..16T}, an H$_{2}$ number density of 9$\times 10^{4}$~cm$^{-3}$ and a kinetic temperature of 50~K are fixed for N113 in our calculations. Assuming a specific CS column density range from 1$\times 10^{12}$~cm$^{-2}$~(\kms)$^{-1}$ to 1$\times 10^{14}$~cm$^{-2}$~(\kms)$^{-1}$, we find that the excitation temperatures vary within 7.4-14.2~K and 8.4-11.2~K for CS (4-3) and CS (5--4), respectively. Based on the modeling results, we also found that the excitation temperatures of different CS rotational transitions vary by less than 3~K (i.e., 30\%). Assuming that CS, C$^{34}$S, and C$^{33}$S share identical collisional rates, we estimate their excitation temperatures by adopting specific column densities 3$\times 10^{12}$~cm$^{-2}$~(\kms)$^{-1}$, 2$\times 10^{11}$~cm$^{-2}$~(\kms)$^{-1}$, and 1$\times 10^{11}$~cm$^{-2}$~(\kms)$^{-1}$ (see Table~\ref{Tab:lin}). Their excitation temperatures are converging to $\sim$7.4~K for the $J$=4-3 transitions and $\sim$8.4~K for the $J$=5-4 transitions, which suggests that molecular excitation should be similar for different CS isotopologues.



Assuming a source size of 40\arcsec\,\citep{2009ApJ...690..580W} and an excitation temperature of 13.1~K, we obtain peak optical depths of $\lesssim$0.08 for both CS (4-3) and CS (5-4). If we consider a relatively low excitation temperature of 5~K, the peak optical depths become $\sim$0.73 and $\sim$0.70 for CS (4-3) and CS (5-4), respectively. On the other hand, the peak intensity ratios between C$^{32}$S and C$^{34}$S are 13.4$\pm$1.6 and 14.8$\pm$0.9 for the $J$=4--3 and $J$=5--4 transitions, respectively. Assuming a $^{32}$S/$^{34}$S isotope ratio of 15 in the LMC \citep{2009ApJ...690..580W}, the peak intensity ratios indicate peak optical depths of $<$0.5 for both CS (4-3) and CS (5-4). Similarly, taking $^{13}$CS (5-4) into account and adopting a $^{12}$C/$^{13}$C isotopic ratio of 50 \citep[e.g.,][]{1999ApJ...512L.143C,1999A&A...344..817H,2009ApJ...690..580W} leads to a peak optical depth of $\sim$0.3 for CS (5--4). These facts suggest that the two CS transitions are likely to have low peak optical depths of $\lesssim$0.7. Given that C$^{34}$S and C$^{33}$S are much less abundant than the main isotopologue by a factor of $>$10 (see discussions below), the C$^{34}$S and C$^{33}$S lines are undoubtedly optically thin.


Under the assumption that all observed transitions are optically thin, we can estimate the LTE column densities of all the CS isotopologues with Eq.~(80) from \citet{2015PASP..127..266M}. The derived column densities are shown in Table~\ref{Tab:lin}, where the associated errors are estimated using a Monte Carlo analysis with 10000 random samples. We also note that if the peak optical depths of the two measured CS transitions were as high as 0.7 (see discussions above), the CS column densities in Table~\ref{Tab:lin} would be underestimated by about 40\%.


Sulfur isotope ratios in the LMC can thus be determined from the column density ratios of pairs of CS isotopologues based on data for the same transitions. A weighted mean value of ratios derived from two transitions is then adopted. All the derived sulfur isotope ratios and lower limits are summarized in Table~\ref{Tab:ratio} together with values derived in a variety of other Galactic environments. Despite the uncertain optical depths of the two CS transitions, these isotope ratios are the most reliable values
for the LMC achieved to date. We note that the newly refined $^{34}$S/$^{33}$S isotope ratio of 2.0$\pm$0.3 strongly contradicts the previously reported lower limit of $\geq$6 for $^{34}$S/$^{33}$S \citep[see Table~8 in ][]{2009ApJ...690..580W}. The ratio is consistent with the ratios observed in ST11 and ST16 on a small scale of $\sim$0.1 pc within their large uncertainties \citep{2020ApJ...891..164S}, whereas our measurements achieve a significantly higher degree of precision. In contrast, our $^{32}$S/$^{34}$S isotope ratio of 15.0$\pm$0.6 appears to be in line with previously estimated values.

 

\begin{table*}[!hbt]
\caption{Sulfur isotope ratios.}\label{Tab:ratio}
\scriptsize
\centering
\begin{tabular}{cccccccccccc}
\hline \hline
                  & \multicolumn{3}{c}{LMC} & \multicolumn{6}{c}{The Milky Way} & \multicolumn{2}{c}{Starburst} \\
\cmidrule(lr){2-4} \cmidrule(lr){5-10} \cmidrule(lr){11-12}
ratio             & N113 \tablefootmark{a}  & ST11\tablefootmark{b} & ST16\tablefootmark{b} & CMZ\tablefootmark{c} & Inner disk\tablefootmark{c}      & Local ISM\tablefootmark{c}  & Outer Galaxy\tablefootmark{c}  & Solar System\tablefootmark{d} & IRC+10216\tablefootmark{e}  & NGC~253\tablefootmark{f} & NGC 4945\tablefootmark{g} \\
\hline
$^{32}$S/$^{34}$S & 15.0$\pm$0.6 & 14$\pm$3 & 17$\pm$2 & 19$\pm$2    &  18$\pm$4 & 24$\pm$4          & 28$\pm$3       & 22.5   & 21.8$\pm$2.6  & $\sim$10 & 13.5$\pm$2.5 \\
$^{32}$S/$^{33}$S & 27.5$\pm$3.5 & 40$\pm$17 & 53$\pm$5 & 70$\pm$16   & 82$\pm$19 & 88$\pm$21        &  105$\pm$19    & 127     & 121$\pm$15    & 50$\pm$30 & ... \\
$^{32}$S/$^{36}$S & $>$81       & ... & ... & 884$\pm$104 & 2382$\pm$368  & 2752$\pm$458 &  4150$\pm$828  & 4748    & 2700$\pm$600    & ... & ... \\
$^{34}$S/$^{33}$S  & 2.0$\pm$0.3 & 3$\pm$2 & 3$\pm$1 & 4.2$\pm$0.2 & 4.3$\pm$0.4  & 4.2$\pm$0.5   &  4.1$\pm$0.3   & 5.6    & 5.6$\pm$0.3   & 5$\pm$3 & ... \\
$^{34}$S/$^{36}$S      & $>$1.3    & ... & ...   & 41$\pm$4    & 122$\pm$18 & 111$\pm$16      &   161$\pm$32   & 200 & 107$\pm$15    & ... & ... \\
\hline
\end{tabular}
\tablefoot{Lower limits of the isotope ratios are derived from the 3$\sigma$ limits in Table~\ref{Tab:lin}. Reference for the sulfur isotopic ratios in different environments: \tablefoottext{a}{This work;} \tablefoottext{b}{\citet{2020ApJ...891..164S};} \tablefootmark{c}{\citet{2023A&A...670A..98Y};} \tablefootmark{d}{\citet{1989GeCoA..53..197A}};
\tablefootmark{e}{\citet{2004A&A...426..219M}}; \tablefootmark{f}{\citet{2021A&A...656A..46M}}; \tablefootmark{g}{\citet{2004A&A...422..883W}}. ``..." indicates that no information is available.
}
\normalsize
\end{table*}

\section{Discussion}\label{Sec:dis}
Table~\ref{Tab:ratio} presents a comparison of sulfur isotope ratios determined in the LMC\footnote{When, in the following, we refer to ``LMC'', we mean specifically the N113 region targeted by us.} and different environments in the Milky Way as well as two nearby starburst galaxies, NGC 253 and NGC 4945. This places our measured ratios of $^{32}$S to $^{33}$S and $^{34}$S for the low-metallicity galaxy into context and promises to reveal clues to their origins. 
The $^{32}$S/$^{34}$S isotope ratio in the LMC appears to be comparable to or slightly lower than different environments in the Milky Way, but higher than those in NGC 4945 \citep[13.5$\pm$2.5,][]{2004A&A...422..883W}, NGC 253 \citep[$\sim$10;][]{2005ApJ...620..210M,2021A&A...656A..46M}, PKS 1830$-$211 \citep[10.5$\pm$0.6;][]{Muller2006}, and B0218+357 \citep[8.1$^{+1.4}_{-1.1}$;][]{Wallstrom2016}. In contrast, our measurements indicate that the $^{32}$S/$^{33}$S and $^{34}$S/$^{33}$S isotope ratios in the LMC are lower than in the Milky Way and likely NGC~253. Even if we increase the CS column densities by 40\% (see the discussions in Sect.~\ref{Sec:res}), the C$^{32}$S/C$^{33}$S isotope ratio would only increase to 39$\pm$5, which suggests that the lower C$^{32}$S/C$^{33}$S isotope ratios in the LMC still hold. On the other side, the $^{32}$S/$^{34}$S isotope ratio derived from CS lines appears to be lower than that obtained from SO lines in NGC~253 \citep{2021A&A...656A..46M}, which indicates that CS transitions in NGC 253 have stronger opacity effects. Considering the uncertainties due to the optical depth effects of the CS transitions, we mainly focus on the $^{34}$S/$^{33}$S isotope ratio and explore various possibilities for an explanation of its origin. 

Isotope ratios can be affected by chemical fractionation. In the case of sulfur, $^{34}$S can be enriched via $^{34}$S$^{+}$+CS$\to$S$^{+}$+C$^{34}$S at low temperatures \citep{2019MNRAS.485.5777L}. However, N113 is found to have kinetic temperatures of $\gtrsim$50~K \citep{2017A&A...600A..16T,2021A&A...655A..12T}, which indicates that chemical fractionation does not play an important role. 

One potential factor influencing sulfur enrichment is the age effect. The $^{34}$S/$^{33}$S isotope ratio would decrease over time, since $^{33}$S appears to be more secondary than $^{34}$S (see below). The lower $^{34}$S/$^{33}$S isotope ratio of $\sim$4.2 in the local ISM compared to the solar value of 5.6 (see Table~\ref{Tab:ratio}) is most likely a result of such an age effect, considering that the Sun was born about 4.6 Gyr ago \citep[e.g.,][]{2010NatGe...3..637B} in the local ISM, which assembled about 7 Gyr ago \citep[e.g.,][]{2001ApJ...554.1044C,2003PASP..115.1187S}. At the Galactocentric distance of $\sim$12~kpc in the outer Galaxy, the stellar metallicity decreases to about 0.5~$Z_{\odot}$ \citep[e.g.,][]{2017A&A...600A..70A}, comparable to that of the LMC \citep[e.g.,][]{1997macl.book.....W}. The comparison between these two regions can eliminate metallicity as a primary factor influencing the observed isotopic differences (the metallicity effect will be discussed in the following). The outer Galaxy should be formed $<$7~Gyr ago, while the LMC was formed $>$10~Gyr ago \citep[e.g.,][]{2009AJ....138.1243H}. Therefore, our observed lower $^{34}$S/$^{33}$S isotope ratio in the LMC compared to the outer Galaxy can also be interpreted as a result of the age effect because of a longer timescale of sulfur enrichment in the LMC.


The LMC's low metallicity could provide another explanation. The observed low $^{14}$N abundances in the LMC have significant implications for the production of $^{18}$O, as $^{18}$O is primarily synthesized through $^{14}$N+$^{4}$He$\to ^{18}$F and subsequent beta decay of $^{18}$F \citep[e.g.,][]{2007hic..book.....C}. Consequently, the low $^{14}$N abundances in the LMC are expected to result in lower stellar yields of $^{18}$O, which is in line with the lower $^{18}$O/$^{17}$O ratios in the LMC than in the Milky Way  \citep[e.g.,][]{1998A&A...332..493H,2009ApJ...690..580W}. Similarly, the alpha capture pathway from $^{14}$N to $^{18}$O, $^{22}$Ne, $^{26}$Mg, $^{30}$Si, and, finally, to $^{34}$S could potentially cause low $^{34}$S/$^{33}$S ratios. However, the efficiency of such an alpha capture sequence is questionable, as previous studies suggest that most of $^{34}$S is predominantly synthesized through oxygen-burning processes \citep[e.g.,][]{1995ApJS..101..181W}. The observed similarity between the $^{18}$O/$^{17}$O isotope ratio \citep[1.7$\pm$0.2;][]{1998A&A...332..493H,2009ApJ...690..580W} and the $^{34}$S/$^{33}$S isotope ratio (2.0$\pm$0.2; this study) appears to be coincidental, because the fusion of oxygen atoms leading to the formation of sulfur occurs in an environment largely deficient in $^{18}$O and $^{17}$O \citep[e.g.,][]{1995ApJS..101..181W}.

Galactic chemical evolution models suggest that the $^{34}$S and $^{33}$S yields are dependent on the metallicity and predict that $^{34}$S/$^{33}$S isotope ratio increases with metallicity \citep[see Fig.~31 in ][]{2020ApJ...900..179K}. This trend can also explain the low $^{34}$S/$^{33}$S isotope ratio in the LMC. On the other hand, 
the Milky Way exhibits decreasing metallicity with increasing Galactocentric distance \citep[e.g.,][]{2011AJ....142..136L}, which would imply a decreasing gradient of the $^{34}$S/$^{33}$S isotope ratios. However, the $^{34}$S/$^{33}$S isotope ratios in the Milky Way are observed to be nearly homogeneous \citep{2023A&A...670A..98Y}. Since the Milky Way is believed to form from the inside out \citep[e.g.,][]{2001ApJ...554.1044C,2012A&A...540A..56P}; this implies an age gradient as a function of Galactocentric distance. As mentioned above, the age effect would cause a decrease of the $^{34}$S/$^{33}$S isotope ratio over time. Thus, the age effect can in principle balance the metallicity dependence of the $^{34}$S/$^{33}$S isotope ratio, giving rise to the nearly homogeneous $^{34}$S/$^{33}$S isotope ratios observed in the Milky Way. 



Another possibility might rest on the star formation history. 
As illustrated by previous studies \citep[e.g.,][]{2018Natur.558..260Z}, the element isotope ratios can be regulated by different stellar populations.
Both $^{33}$S and $^{34}$S are not purely primary isotopes because they can be produced by oxygen burning and explosive nucleosynthesis of massive stars as well as neutron capture by $^{32}$S and $^{33}$S \citep[e.g.,][]{2007hic..book.....C}. Their isotope ratios are thus based on their respective yields from these processes. Previous studies suggest that core-collapse supernovae of massive stars can produce $^{34}$S more efficiently than $^{33}$S \citep[e.g.,][]{1995ApJS..101..181W,2008MNRAS.390.1710H,2020ApJ...900..179K}. On the other hand, novae originating from low- and intermediate-mass stars can slightly overproduce $^{33}$S because of proton captures \citep[e.g.,][]{1998ApJ...494..680J}. Hence, the observed low value can in principle be caused by a larger population of low- and intermediate-mass stars or a smaller population of massive stars.
We note however, that the $^{32}$S/$^{33}$S ratio of 28 we determine for the LMC is much lower than the value of 121 that was derived by \citet{2004A&A...426..219M} for the archetypal carbon-rich AGB object IRC+10216 (also known as CW Leo; see Table~\ref{Tab:ratio}), which (by definition) is a low-mass star. Their $^{32}$S/$^{34}$S ratios are comparable: 15 (LMC) vs. 22 (IRC+10216). In the former scenario, we would anticipate relatively high nitrogen abundances in the LMC, because $^{14}$N is primarily synthesized in the CN cycle of low- and intermediate-mass stars \citep[e.g.,][]{1994ARA&A..32..191W,2007hic..book.....C,2022A&ARv..30....7R}. However, this scenario contradicts the fact that the LMC exhibits a rather low nitrogen abundance \citep{1982ApJ...252..461D,2016ApJ...818..161N}, which is typical for low-metallicity galaxies \citep{2011A&A...529A.149G}. The latter hypothesis seems to conflict with the measured Salpeter-like or top-heavy initial mass function (IMF) in the LMC \citep[e.g.,][]{1995ApJ...438..188M,2012A&A...547A..23B,2018Sci...359...69S} but this contradiction is not definitive. The inconsistency could be attributed to the variable star formation rate in the LMC \citep[e.g.,][]{2009AJ....138.1243H} and the fact that the measured isotope ratios reflect cumulative sulfur enrichment over time. The measured IMF likely represents massive stars formed only tens of millions of years ago \citep[e.g.,][]{1995ApJ...438..188M}, while sulfur enrichment could extend back to the formation of the earliest generation of stars at least several billion years ago. In other words, sulfur isotope ratios can be influenced by potential variations of the IMF as a function of the star formation history. The star formation history of the LMC indicates a quiescent epoch spanning roughly 5--12 Gyr ago when star formation rates were quite low \citep{2009AJ....138.1243H}. The low number of massive stars during this period resulted in lower production of $^{34}$S in the LMC and in N113 in particular. Therefore, the low $^{34}$S/$^{33}$S isotope ratio may also be attributed to the star formation history.

In light of these findings, we propose that the observed low $^{34}$S/$^{33}$S isotope ratio in N113 could be a consequence of the combination of the age effect, low metallicity, and star formation history. We note that our measurements solely provide constraints on the sulfur isotope ratios within N113. The metallicity does not vary much across the LMC~\citep[0.3--0.5$Z_{\odot}$, e.g.,][]{1997macl.book.....W,2002A&A...396...53R}, indicating that the ISM is well mixed. This is further supported by the nearly constant $^{18}$O/$^{17}$O isotopic ratio in different regions \citep{1998A&A...332..493H,2009ApJ...690..580W} as well as the reasonably consistent $^{34}$S/$^{33}$S isotope ratios observed in N113, ST11, and ST16. Consequently, our measured $^{34}$S/$^{33}$S isotope ratio could conceivably represent a typical value for the LMC. Nevertheless, the degree to which this value faithfully traces the sulfur isotopic ratios prevailing throughout the LMC remains a topic of future studies. To gain a deeper insight into the sulfur enrichment of the whole galaxy, additional measurements of the sulfur isotopic ratios toward different parts of the LMC are mandatory.


\section{Summary and conclusions}\label{Sec:sum}
In this study, we carried out pointed observations toward N113 in the Large Magellanic Cloud (LMC) using the APEX telescope. Our observations have led to the unambiguous detection of the C$^{33}$S (4-3) and (5--4) lines, providing conclusive evidence of the existence of C$^{33}$S on a scale of $\sim$5~pc in the LMC. Measurements of optically thin tracers allow us to determine the $^{34}$S/$^{33}$S isotope ratio, yielding a value of 2.0$\pm$0.3. This determination represents the most reliable $^{34}$S/$^{33}$S isotope ratio measurement on a cloud scale in a low-metallicity galaxy. Notably, the measured $^{32}$S/$^{33}$S and $^{34}$S/$^{33}$S isotope ratios in the LMC are about a factor of two lower than the corresponding values observed in the Milky Way. We postulate that this discrepancy can be attributed to a combination of the age effect, low metallicity, and star formation history. 

\section*{ACKNOWLEDGMENTS}\label{sec.ack}
The data was collected under the Atacama Pathfinder EXperiment (APEX) project, led by the Max Planck Institute for Radio Astronomy under the umbrella of the ESO La Silla Paranal Observatory. We acknowledge the APEX staff for their assistance with our observations. C.H. acknowledges support by Chinese Academy of Sciences President’s International Fellowship Initiative under grant No. 2023VMA0031. C.-H.R. C. acknowledges support from the Deutsches Zentrum f{\"u}r Luft- und Raumfahrt (DLR) grant NS1 under contract no. 50 OR 2214. X.D.T. acknowledges the support of the Chinese Academy of Sciences (CAS) ``Light of West China" Program under grant No. xbzg-zdsys-202212, 
the Tianshan Talent Program of Xinjiang Uygur Autonomous Region under grant No. 2022TSYCLJ0005, and the Natural Science Foundation of Xinjiang Uygur Autonomous Region under grant No. 2022D01E06. This research has made use of NASA's Astrophysics Data System. We thank our referee Sergio Martin for his valuable comments that have contributed to the improvement of this paper. We also thank Takashi Shimonishi for his comments on this draft.


\bibliographystyle{aa}
\bibliography{references}

\begin{appendix}

\end{appendix}

\end{document}